\documentstyle[12pt,aasms4]{article}

\begin{document}

\title{Neutrino-Driven Jets and Rapid-Process Nucleosynthesis}
\author{Shigehiro Nagataki}
\noindent
{Department of Physics, School of Science, the University
of Tokyo, 7-3-1 Hongo, Bunkyoku, Tokyo 113, Japan}

\begin{abstract}
We have studied whether the jet in a collapse-driven
supernova can be a key process for the rapid-process (r-process)
nucleosynthesis.
We have examined the features of a steady, subsonic, and rigidly
rotating jet in which the centrifugal force is balanced by the
magnetic force.
As for the models in which the magnetic field is weak and angular
velocity is small, we found that the r-process does not occur
because the final temperature is kept to be too high and the
dynamical timescale becomes too long when the neutrino
luminosities are set to be high.
Even if the luminosities of the neutrinos are set to be low, which
results in the low final temperature, we found that the models do
not give a required condition to produce the r-process matter.
Furthermore, the amount of the mass outflow seems to be too little
to explain the solar-system abundance ratio in such low-luminosity
models. As for the models in which the magnetic field
is strong and angular velocity is large,
we found that the entropy
per baryon becomes too small and the dynamical timescale becomes too long.
This tendency is, of course, a bad one for the production of the
r-process nuclei. As a conclusion, we have to say that it is difficult
to cause a successful r-process nucleosynthesis in the jet models
in this study.
\end{abstract}
\keywords{nucleosynthesis, abundances --- stars: magnetic --- stars: rotation
--- supernovae: general}

\section{INTRODUCTION} \label{intro}
\indent

It is one of the most important astrophysical problems that the sites where
the rapid-process (r-process) nucleosyntesis occurs are not still
known exactly. There are, at least, three reasons that make the study
on r-process nucleosynthesis important. One of them is a very pure
scientific interest. The mass numbers of the products of r-process
nucleosynthesis are very high ($A$ = 80--250), which means that the
most massive nuclei in the universe are synthesized through the r-process. 
You can guess easily that the situation in which the r-process nuclei
are synthesized is a very peculiar one in the universe. We want to know
where, when, and how the r-process nuclei are formed. Second reason is
that some r-process nuclei can be used as chronometers. For example,
the half-lives of $\rm ^{232}Th$ and $\rm ^{238}U$ are 1.405$\times 10^{10}$
yr and 4.468$\times 10^{9}$ yr, respectively. So if we can
predict the mass-spectrum of the products of r-process nucleosynthesis
precisely, we can estimate the ages of metal-poor objects which
contain the r-process nuclei by observing its abundance ratio.
Third reason is that some r-process nuclei can be used as tools of
the study on the chemical evolution in our Galaxy (e.g., \cite{ishimaru99}),
which has a potential to reveal the history of the evolution of our
Galaxy itself. Due to the reasons mentioned above, the study on the
r-process nucleosynthesis is very important.

The conditions in which the r-process nucleosynthesis occurs successfully
are (e.g., \cite{hoffman97}): (i) neutron-rich ($n_n \ge \rm 10^{20}
\; cm^{-3}$), (ii) high entropy per baryon, (iii) small dynamical
timescale, and (iv) small $Y_e$. This is because r-process nuclei are
synthesized throuth
the non-equilibrium process of the rapid neutron capture on the seed
nuclei (iron-group elements). In other words, a highly dense, explosive,
and neutron--rich site with high entropy will be a candidate for the location
where the r-process nucleosynthesis occurs.

The most probable candidates for the sites
are collapse-driven supernovae (e.g., \cite{woosley94}) and/or neutron star
mergers (e.g., \cite{freiburghaus99}). This is because these
candidates are thought
to have a potential to meet the requirements mentioned above.
However, we think that the collapse-driven supernovae are thought to
be more probable sites than the neutron star mergers, because metal
poor stars already contain the r-process nuclei (e.g., \cite{freiburghaus99}).
In fact, Mcwilliam et al. (1995) reported that the abundance of Eu
can be estimated in 11 stars out of 33 metal-poor stars. These
observations prove that r-process nuclei are produced
from the early stage
of the star formation in our Galaxy. Taking the event rate of
collapse-driven supernovae ($10^{-2} \; \rm yr^{-1} \; Gal^{-1}$;
\cite{vandenbergh91}) and neutron star merger ($10^{-5} \; \rm yr^{-1}
\; Gal^{-1}$; \cite{vanden96}; \cite{bethe98}) into consideration,
collapse-driven supernovae
are favored since they can supply the r-process nuclei from the early stage
of the star formation in our Galaxy. Also, Cowan et al (1999) reported
that the abundance ratio of r-process nuclei in metal poor stars
are very similar to that in the solar system. This proves that r-process
nuclei are synthesized through the similar conditions. This will be
translated that, at least, most of the r-process nuclei are from one
candidate. So we assume in this paper that most of the r-process nuclei
are synthesized in the collapse-driven supernovae.

There are many excellent and precise analytic and/or numerical
calculations on the r-process nucleosynthesis in the collapse-driven
supernovae.
However, it would be able to be said that there is no report that
the r-process nuclei can be reproduced completely.
For example, Takahashi et al. (1994) performed numerical simulations
assuming Newtonian gravity and reported that entropy per baryon in the
hot bubble is about 5 times smaller than the required value.
Qian and Woosley (1996; hereafter QW96) also reported analytic treatments of
the neutrino-driven winds from the surface of the proto-neutron star.
At the same time, their analytical treatments are tested and confirmed
by numerical methods. However, the derived entropy by their wind solutions
are shown to fall short, by a factor of 2--3, of the value required
to produce a strong r-process (\cite{hoffman97}).
Otsuki et al. (2000) have surveyed the effects
of general relativity parametrically. They reported that r-process
can occur in the strong neutrino-driven winds ($L_{\nu} \sim 10^{52}$
erg $\rm s^{-1}$) as long as a massive ($\sim$2.0 $M_{\odot}$) and
compact ($\sim$ 10 km) proto-neutron star is formed.
It is very interesting because such
a solution can not be found in the frame work of Newtonian gravity
(\cite{qian96}). Such a solution is confirmed
by the excellent numerical calculations (\cite{sumiyoshi99}).
However, the equation
of state (EOS) of the nuclear matter has to be too soft to achieve such
conditions. In fact, the r-process nuclei can not be
produced in the numerical simulations with an normal EOS
(\cite{sumiyoshi99}).

There is only one report that r-process nucleosynthesis occurred
successfully. That is the work done by Woosley et al
(1994; here after WWMHM94).
In their numerical simulation, the entropy per baryon becomes higher
and higher as the time goes on. Finally, at very late phase of
neutrino-driven wind ($\sim 10$ s after the core-collapse), successful
r-process occurs. However, there are some questions on their results.
One is that the reason why the entropy per baryon becomes high at the
late phase is unclear. In fact, when we adopt the analytic formulation
of QW96, it is concluded that such a high entropy should not be obtained.
It is true that general relativistic effects are included in WWMHM94,
but such a high entropy could not be obtained in the work done
by Otsuki et al (2000). So the discrepancy between WWMHM94 and QW96
can not be explained by the general relativistic effects alone.
Also, WWMHM94 has a problem that the nuclei whose mass numbers are
$\sim$ 90 are much produced in the early stage of the neutrino-driven
winds. So if we try to reproduce the mass-spectrum of the solar system
abundances, we have to abandon the matter which is made at the early
stage of the neutrino-driven winds.
Even worse, the successful mass-spectrum
at the late phase of neutrino-driven winds is destroyed when the reactions of
neutral-current neutrino spallations of nucleons from $\rm ^{4}He$ are
taken into consideration (\cite{meyer95}). In their paper, it is
reported that (30--50)$\%$ increase in the entropy is needed in order
to restore the $A$ = 195 peak, which is extremely large modifications to the
model. So, although the work done by WWMHM94 is the very remarkable and
interesting one, it would not be concluded that the problem of r-process
nucleosynthesis has been solved completely.

Due to the reason mentioned above, it will be natural to think
that there may be an (some) effect(s) that
will help the r-process nucleosynthesis. There are some papers that
mention the effects of jet (\cite{symbalisty84}; \cite{shimizu94};
\cite{nagataki00}) that is generated around the polar region
of the highly rotating (the period $T\sim$1ms) and/or
magnetized ($\sim 10^{15}$ G) proto-neutron star (\cite{leblanc70};
\cite{symbalisty84}; \cite{shimizu94}; \cite{yamada94}; \cite{fryer99}).
Since the physical conditions in the jet are different from those
in a spherical explosion, the products of the nucleosynthesis are expected
to be different. In fact, it is reported that the products of the explosive
nucleosynthesis in a jet-like explosion in Si-- and O-- rich layer are
much different from those in a spherical explosion (\cite{nagataki97};
\cite{nagataki00}). So it is natural to think that the products of explosive
nucleosynthesis in the Fe core (that is, in the hot bubble) will
be also changed due to the effects of the jet.
This is the reason why we should examine whether the effects of the
jet can help the synthesis of r-process nuclei or not.

Since there is no numerical simulation on the r-process nucleosynthesis
in the jet during the neutrino-driven wind phase, our final goal is to
perform such realistic numerical simulations. However, such a numerical
simulation will be
a heavy task. We also think that the results of such numerical simulations
will not be understood completely without analytical studies.
So, before performing such numerical simulations, we examine the physical
conditions of the jet using a simple model. In this paper, we study a steady,
subsonic, and rigidly
rotating flow of the jet that is driven by neutrinos. 
We study the effects of the jet on the physical conditions such as the
entropy per baryon and the dynamical timescale. Finally, we discuss
whether the effects of the jet can help the synthesis of r-process nuclei.

In section~\ref{model}, we explain the formulation for the jet in the
hot bubble. In section~\ref{result}, we show the results.
Summary and discussions are presented in section~\ref{summary}.

\section{FORMULATION OF JET} \label{model}
\subsection{Basic Equations}\label{basic}
\indent

In Gaussian units, the Euler's equation acted on by electromagnetic forces
can be written as (\cite{shapiro83})
\begin{eqnarray}
v \frac{dv}{dt} = - \frac{1}{\rho} \nabla P - \nabla \Phi 
-\frac{1}{8 \pi \rho} \nabla B^2 + \frac{1}{4 \pi \rho} (\vec{B} \cdot \nabla)
\vec{B}. 
\label{eqn1}
\end{eqnarray}
Here
\begin{eqnarray}
\frac{d}{dt} = \frac{\partial}{\partial t} + \vec{v} \cdot \nabla
\label{eqn2}
\end{eqnarray}
is the Lagrangian time derivative following a fluid element.

In this paper, we study a steady jet which has $\phi$-symmetry
around the polar region of the proto-neutron star.
So we use the cylindrical coordinate ($r$, $\phi$, $z$) for convenience.
The origin $z$=0 is set at the center of the proto-neutron star.
In the cylindrical coordinate, the Euler equation can be written as
\begin{eqnarray}
\left(
\begin{array}{c}
v_r \frac{\partial v_r}{\partial r}   + v_z \frac{\partial v_r}{\partial z}
- \frac{v^2_{\phi}}{r} \\
v_r \frac{\partial v_\phi}{\partial r}+ v_z \frac{\partial v_\phi}{\partial z}
+ \frac{v_r v_\phi}{r} \\
v_r \frac{\partial v_z}{\partial r} + v_z \frac{\partial v_z}{\partial z}
\end{array}{}
\right) =
- \frac{1}{\rho} 
\left(
\begin{array}{c}
\frac{\partial P}{\partial r} \\
0 \\
\frac{\partial P}{\partial z} 
\end{array}{}
\right) 
- 
\left(
\begin{array}{c}
0 \\
0 \\
\frac{G M}{z^2}
\end{array}{}
\right)
+ \frac{1}{4 \pi \rho}
\left(
\begin{array}{c}
\left( \frac{\partial B_r}{\partial z} - \frac{\partial B_z}{\partial r}
\right)B_z  - \frac{B_\phi}{r}\frac{\partial}{\partial r} (r B_\phi) \\
\frac{B_r}{r} \frac{\partial}{\partial r} (rB_\phi) + B_z
\frac{\partial B_\phi}{\partial z} \\
- B_r \left( \frac{\partial B_r}{\partial z} - \frac{\partial B_z}{\partial r}
\right) - B_\phi \frac{\partial B_\phi}{\partial z}
\end{array}{}
\right).
\label{eqn3}
\end{eqnarray}

Conservation of mass requires that
\begin{eqnarray}
\dot{M} = \int_0^D  2 \pi r \rho v_z dr,
\label{eqn4}
\end{eqnarray}
where $\dot{M}$ and $D$ are the constant mass outflow rate in the ejecta and
the radius of the jet, respectively.

The equation for the evolution of material energy, $\epsilon$, is
\begin{eqnarray}
\rho \dot{q} &=& \nabla \cdot (\rho \epsilon \vec{v}) 
+ P \nabla \cdot \vec{v} \\ 
             &=& \left(  \frac{1}{r} \frac{\partial}{\partial r}
(r \rho \epsilon v_r) + \frac{\partial}{\partial z} 
(\rho \epsilon v_z)      \right)  +
P \left( \frac{1}{r} \frac{\partial}{\partial r} (r v_r) +
\frac{\partial}{\partial z} v_z
\right)
\label{eqn5}
\end{eqnarray}
where $\dot{q}$ is
the net specific heating rate due to neutrino interactions (\cite{qian96}).
In this study, we consider three neutrino heating and/or cooling processes
(neutrino absorption on free nucleons, neutrino scattering processes
on the electrons and positrons, and electron capture on free nucleons)
as
\begin{eqnarray}
\dot{q} = \dot{q}_{\nu N} + \dot{q}_{\nu e} - \dot{q}_{e N},
\label{eqn51}
\end{eqnarray}
where
\begin{eqnarray}
\dot{q}_{\nu N} = 1.55 \times 10^{-5}  N_{\rm A} \left [(1-Y_e)L_{\nu_e,51}
\epsilon^2_{\nu_e, \rm MeV} + Y_e L_{\bar{\nu}_e,51} 
\epsilon^2_{\bar{\nu}_e, \rm MeV}      \right] \frac{1 - x}{R_6^2} \;\;\; 
\rm  \left [ erg \; s^{-1} \; g^{-1} \right] ,
\label{eqn52}
\end{eqnarray}
\begin{eqnarray}
\dot{q}_{\nu e} = 3.48 \times 10^{-6} N_{\rm A} \frac{T^4_{\rm MeV}}{\rho_8}
\left ( 
L_{\nu_e,51} \epsilon_{\nu_e, \rm MeV} +
L_{\bar{\nu}_e,51} \epsilon_{\bar{\nu}_e, \rm MeV} +
\frac{6}{7}L_{\nu_{\nu},51} \epsilon_{\nu_{\nu}, \rm MeV}       
\right ) 
\frac{1-x}{R_6^2} \;\;\; 
\rm \left [  erg \; s^{-1} \; g^{-1}  \right ] ,
\label{eqn53}
\end{eqnarray}
and
\begin{eqnarray}
\dot{q}_{e N} = 3.63 \times 10^{-6} N_{\rm A} T_{\rm MeV}^6 \;\;\; 
\rm \left [  erg \; s^{-1} \; g^{-1} \right ]  .
\label{eqn54}
\end{eqnarray}
Here $R_6$ is the neutrino sphere radius in units of $10^6$ cm,
$x = (1 - R^2/z^2)^{1/2}$, $N_{\rm A}$ is Avogadro's number, $L_{\nu, 51}$
is the individual neutrino luminosity in $10^{51}$ erg $\rm s^{-1}$,
$\epsilon_{\nu, \rm MeV}$ is an appropriate neutrino energy $\epsilon_{\nu}$
in MeV (\cite{qian96}). In this study, we set $\dot{q}$=0 at
$T \le $0.5 MeV, because free nucleons are bound into $\alpha$-particles
and heavier nuclei and electron-positron pairs annihilate into photons.

The pressure $P$ and internal energy $\epsilon$ are determined approximately
by the relativistic electrons and positrons and photon radiation as long as
$T \ge 0.5$ MeV. So the pressure and internal energy can be written as
\begin{eqnarray}
P = \frac{11 \pi ^2}{180}\frac{k^4T^4}{\hbar^3   c^3} \;\;\; \rm \left [
dyn \; cm^{-2} \right ]
\label{eqn6}
\end{eqnarray}
and
\begin{eqnarray}
\epsilon = \frac{11 \pi ^2}{60}\frac{k^4T^4}{\hbar^3 c^3 \rho} \;\;\; \rm
\left [  erg \; g ^{-1}        \right ],
\label{eqn7}
\end{eqnarray}
where $k$ and $\hbar$ are Boltzmann and Planck constants, respectively.
These are the basic equations in this paper.
Precisely, we should consider the effects of annihilations of
electron-positron pairs on the dynamics at $T \le 0.5$ MeV.
However, its effects seems to be little (\cite{sumiyoshi99}).

\subsection{Model for the Jet}\label{jetmodel}
\indent

In this study, we try to construct a simple steady state for a jet
so that we can understand the effects of magnetic fields clearly.
So we assume in this study that the form of the magnetic fields is
\begin{eqnarray}
\vec{B} = (0,0,B_\phi (r,z)).
\label{eqn8}
\end{eqnarray}
Next, we seek a solution for the jet rotating rigidly. That is,
we seek a solution $\vec{v}$ =
(0,$\Omega r$,$v_z(r,z)$) at $z \ge R$, where $\Omega$ is the angular velocity
of the proto-neutron star. From Eq.~(\ref{eqn3}), $\vec{B}$ and $\vec{v}$
must meet the following relation:
\begin{eqnarray}
\left(
\begin{array}{c}
- \Omega^2 r \\
v_z \frac{\partial v_z}{\partial z} 
\end{array}{}
\right) =
- \frac{1}{\rho}
\left(
\begin{array}{c}
\frac{\partial P}{\partial r}\\
\frac{\partial P}{\partial z}
\end{array}{}
\right) 
-
\left(
\begin{array}{c}
0 \\
\frac{G M}{z^2}
\end{array}{}
\right) 
- \frac{1}{4 \pi \rho}
\left(
\begin{array}{c}
\frac{B_\phi}{r}\frac{\partial}{\partial r} (r B_\phi) \\
B_\phi \frac{\partial B_\phi}{\partial z}
\end{array}{}
\right).
\label{eqn9}
\end{eqnarray}
In this case, there is a useful approximate solution. That is,
\begin{eqnarray}
B_\phi = (2 \pi \rho)^{\frac{1}{2}} \Omega r,
\label{eqn10}
\end{eqnarray}
where $\rho$ and $P$ are assumed to be a function of $z$ alone.
It is noted that this approximate solution is valid as long as $\Omega D$
is small enough. In fact, the derivatives of $\rho$ and $T$ by $z$ can
be written as
\begin{eqnarray}
\frac{d \rho}{dz} = - \frac{\frac{P  \dot{q}}{v_z \epsilon} + 
\frac{\rho GM}{z^2}}{\frac{P}{\epsilon \rho} 
\left( \epsilon + \frac{P}{\rho}   \right) - v_z^2 + \frac{(\Omega r)^2}{4}}
\label{eqn11}
\end{eqnarray}
\begin{eqnarray}
\frac{d T}{dz} = \frac{\frac{\dot{q}}{v_z} + \left( \frac{\epsilon}{\rho} 
+ \frac{P}{\rho^2}  \right)\frac{d \rho}{d z}   }{\frac{4 \epsilon}{T}}.
\label{eqn12}
\end{eqnarray}
It is clearly understood from Eq.~(\ref{eqn11}) that the
solution holds valid as long as $\Omega D$ is small enough.
In this case, Eq.~(\ref{eqn4}) can be rewritten as
\begin{eqnarray}
\dot{M} = \pi D^2 \rho v_z.
\label{eqn121}
\end{eqnarray}

From Eqs.~(\ref{eqn9}) and~(\ref{eqn10}), we can see that
the magnetic fields accelerate
the matter as long as the density gradient is negative.
We are also able to see clearly from Eq.~(\ref{eqn10})
that small $\Omega D$ corresponds
to weak magnetic field. So we call such solutions weak field
solutions in this
paper. On the other hand, we also want to see what can happen as the
magnetic fields
are getting stronger and stronger. So, although the truth is that these
solutions are inapplicable to the case of strong magnetic fields, we
extend these solutions to the case in which $\Omega D$ is relatively large.
By so doing, we will be able to find at least what happens in
the case of strong
magnetic fields qualitatively. We call such solutions strong
field solutions in this paper. In section~\ref{result}, we show the
results for the weak and strong field solutions, respectively.

\subsection{Boundary Conditions}\label{boundary}
\indent

In this study, the surface of the proto-neutron star is considered
as the inner boundary.
The inner boundary conditions are composed of density, luminosities of
neutrinos, mass and radius of the proto-neutron star, velocity of the
outflow, and $\Omega D$.
Initial temperature and final electron fraction are determined by
these parameters as (\cite{qian96})
\begin{eqnarray}
T_{i} = 1.19 \times 10^{10} \left[ 1 + 
\frac{L_{\nu_{e}}}{L_{\bar{\nu}_{e}}}
\left( \frac{\epsilon_{\nu_{e}, \rm MeV}}{\epsilon_{\bar{\nu}_{e}, \rm MeV}}
\right)^2
\right]^{\frac{1}{6}} L^{\frac{1}{6}}_{\bar{\nu}_e, \rm 51}
R^{-\frac{1}{3}}_{\rm 6} \epsilon^{\frac{1}{3}}_{\bar{\nu}_e, \rm MeV} \;\;\;
\rm \left [  K     \right ]
\label{eqn13}
\end{eqnarray}
and
\begin{eqnarray}
Y_{e,f} = \left( 1 + \frac{L_{\bar{\nu}_{e}}}{L_{\nu_{e}}}
\frac{\epsilon_{\bar{\nu}_{e}, \rm MeV} - 2\Delta + 
1.2 \Delta^2/ \epsilon_{\bar{\nu}_{e}, \rm MeV}  }{\epsilon_{\nu_{e}, \rm MeV} + 2\Delta +
1.2 \Delta^2/\epsilon_{\nu_{e}, \rm MeV}}          \right)^{-1}, 
\label{eqn14}
\end{eqnarray}
where $L_{\nu, \rm 51}$ is the individual neutrino luminosity in $10^{51}$
$\rm ergs \; s^{-1}$, $R_{6}$ is the neutron star radius in $10^{6}$ cm,
$\Delta$ = 1.293 MeV is the neutron-proton mass difference,
and $\epsilon_{\nu, \rm MeV}$ is a neutrino energy in MeV. We assume that
the neutron star radius is equal to the neutrino sphere radius.

In this study, the luminosities of neutrinos are assumed to be common
(\cite{qian96}; \cite{otsuki00}). The energy of neutrinos are assumed
to be 12, 22, and 34 MeV for $\nu_{e}$, $\bar{\nu}_{e}$, and other neutrinos,
respectively (\cite{woosley94}; \cite{qian96}; \cite{otsuki00}).
Surface density is assumed to be $10^{10}$ $\rm g \; cm^{-3}$
(\cite{otsuki00}).
Initial velocity of the outflow is chosen so that $\dot{M}$ becomes less than
$\dot{M}_{\rm crit}$, where $\dot{M}_{\rm crit}$ is the critical value for
supersonic solution. We have to emphasize that the flow
is assumed to be subsonic and contain no critical point.
Previous works also adopted this assumption (\cite{qian96}; \cite{otsuki00}).
In case we try to survey the flow that contains
a transition point like a shock front,
we can not use Eqs.~(\ref{eqn5}),~(\ref{eqn9}),~(\ref{eqn11}),
and~(\ref{eqn12}).
This is because these differential equations diverge and break down.
In order to treat such a flow that contains a discontinuity,
we have to use the Rankine-Hugoniot relation instead of these equations.
We will examine such flows in the forthcoming paper.
$D$ is assumed to be $10^5$ cm,
which is about one tenth of the proto-neutron star radius.
We note that an appropriate environment can exist around the jet,
that is, there is a solution that meets Eq.~(\ref{eqn9})
for $r \ge D$ in which $B_{\phi}$ and $v_{\phi}$
becomes zero at $r \rightarrow \infty$.
So the rigid rotating flow at $r \le D$ is not 
an unrealistic solution at all. In this study,
we consider only the region $r \le D$ in
which the matter rotates rigidly for simplicity.
Other parameters are changed parametrically.
The parameters employed as well as the model names are given
in Table~\ref{tab1}. We take $z$ = $10^9$ cm for the radius of the
outer boundary (\cite{otsuki00}). Since the radius of the Fe core
is about $10^8$ cm, we think the radius of the outer boundary is
large enough to investigate the r-process nucleosynthesis that
occurs in the hot bubble.

\placetable{tab1}

\section{RESULTS} \label{result}
\indent
\subsection{Weak Field Solutions}\label{weak}
\indent

Output parameters are shown in Table~\ref{tab2}. Entropy per baryon ($S$),
dynamical timescale ($\tau_{\rm dyn}$), analytically estimated dynamical
timescale ($\tau_{\rm dyn.ana}$), electron fraction ($Y_e$),
and temperature ($T$) at the outer
boundary ($z$ = $10^9$ cm) are shown in the table.
Entropy per baryon in radiation dominated gas can be written as
\begin{eqnarray}
S/k = \frac{11 \pi^2}{45} \frac{k^3}{\hbar^3 c^3} \frac{T^3}{\rho/m_N}
\label{eqn15}
\end{eqnarray}
as long as $T \ge$ 0.5 MeV. $m_N$ is the nucleon rest mass.
Even if the annihilation of the electron-positron pairs occurs,
the entropy per baryon is conserved.
The definition of the dynamical timescale is
the time for the temperature to decrease from 0.5 MeV to 0.2 MeV.
This is because r-process nucleosynthesis occurs in this temperature
range (\cite{woosley94}; \cite{takahashi94}; \cite{qian96}).
The reason why the dynamical timescales
are not written in some models is that temperature does not decrease
to 0.2 MeV within $z$ = $10^9$ cm. The definition of the analytically
estimated dynamical timescale is described later (see Eq.~(\ref{eqn25})).

\placetable{tab2}

\begin{figure}
\epsscale{1.0}
\plotone{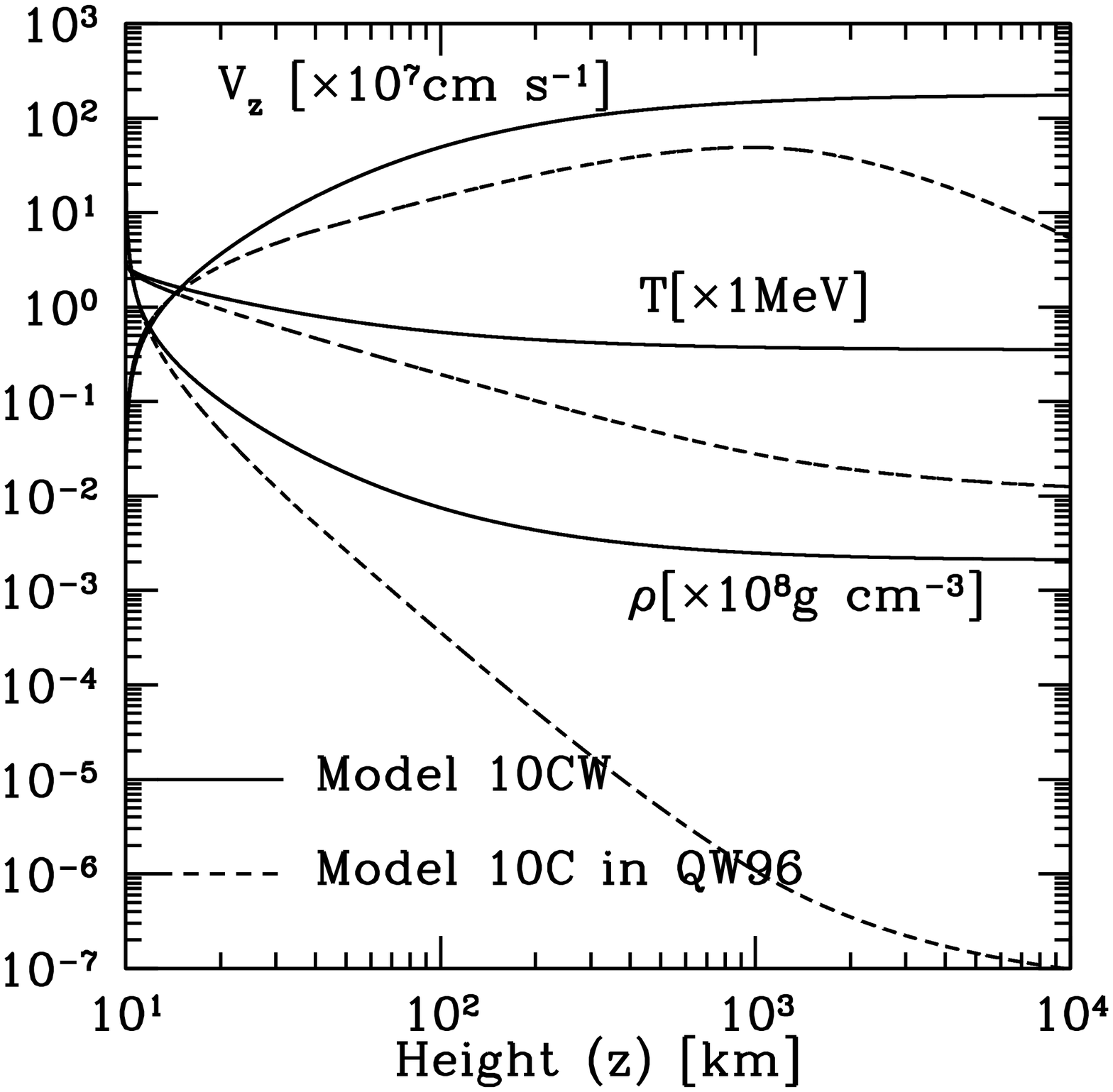}
\figcaption{
Outflow velocity, temperaure, and density as a function of height ($z$)
from the center of the proto-neutron star. These quanta are written
in unit of $10^7$ cm $\rm s^{-1}$, 1 MeV, and $10^8$ g cm$^{-3}$,
respectively. Solid lines correspond to Model 10CW, whereas dashed
lines correspond to Model 10C in QW96.
\label{fig1}}
\end{figure}

\begin{figure}
\epsscale{1.0}
\plotone{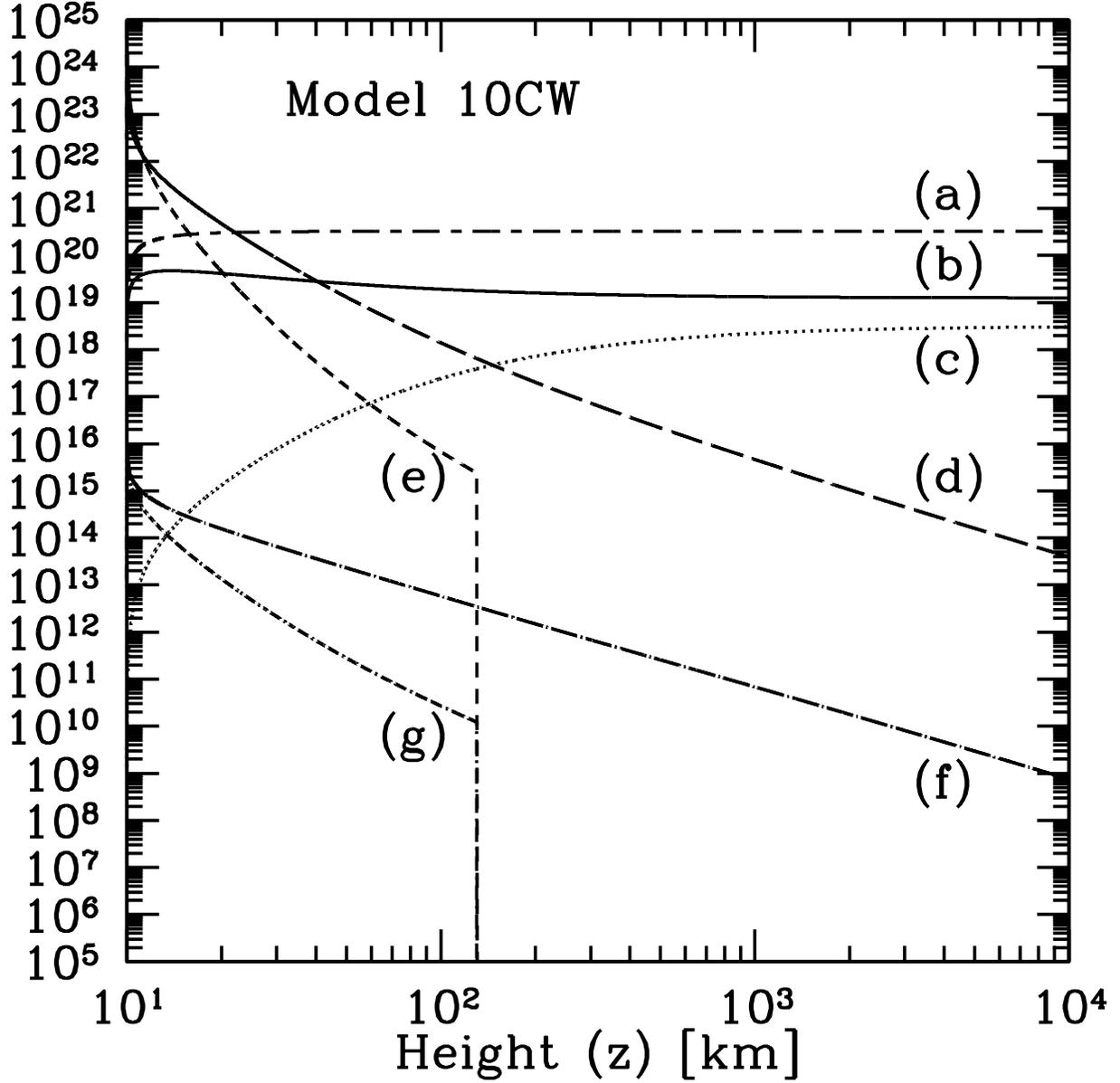}
\figcaption{
Absolute values (in cgs units)
of the components of Eqs.~(16) and~(17) for Model 10CW.
Lines (a)-(g) correspond to $4\epsilon/T$,
$(P/\epsilon \rho)(\epsilon+P/ \rho)$, $v_z^2$, $\rho GM/z^2$,
$P\dot{q}/v_z \epsilon$, $(\epsilon/\rho + P/\rho^2)\frac{d \rho}{d z}$,
and $\dot{q}/v_z$ as a function of $z$, respectively.
The discontinuities of lines (e) and (g) at $z \sim  10^7$ cm
reflect the freezeout of the neutrino reactions (see subsection 2.1).
\label{fig2}}
\end{figure}

\begin{figure}
\epsscale{1.0}
\plotone{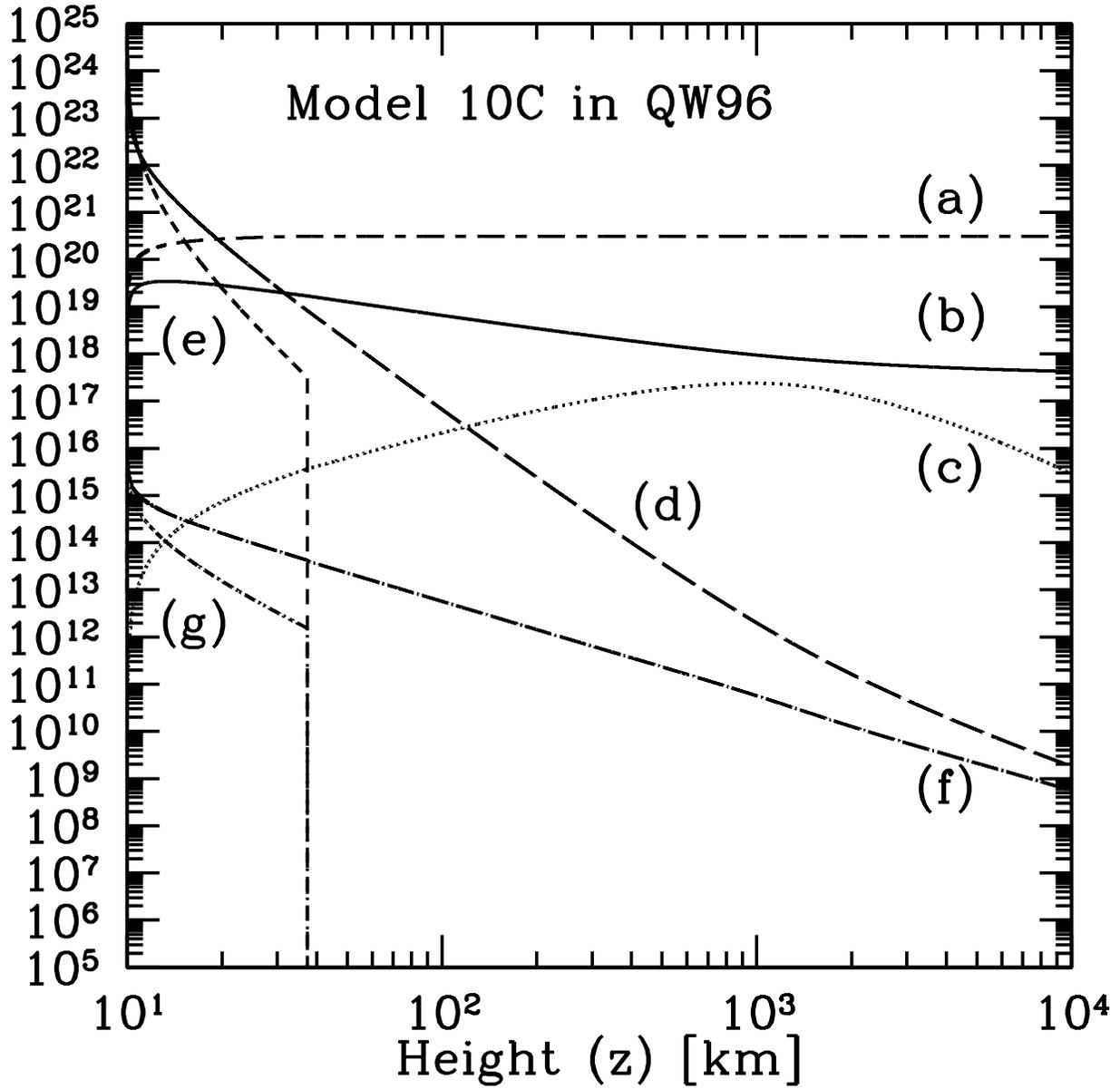}
\figcaption{
Same as Figure 2, but for Model 10C in QW96.
\label{fig3}}
\end{figure}

\begin{figure}
\epsscale{1.0}
\plotone{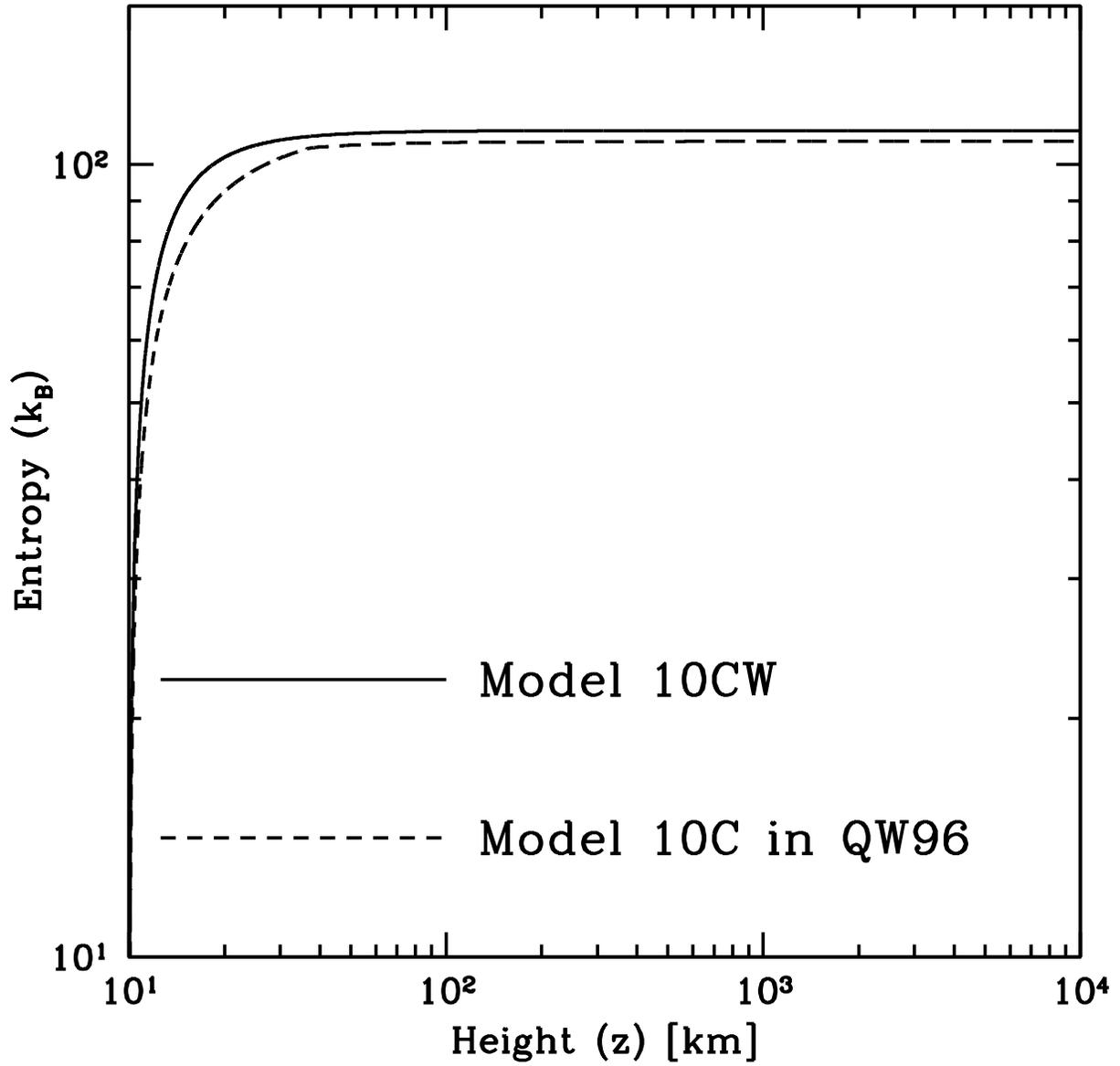}
\figcaption{
Entropy per baryon as a function of height ($z$)
from the center of the proto-neutron star. Solid line corresponds to that
of Model 10CW. Sort-dashed line corresponds to that of Model 10C in QW96.
\label{fig4}}
\end{figure}

There is a tendency that $S$ and $T_b$ in the models studied here
are higher than those in the models of QW96. For example,
we show in Figure~\ref{fig1} the outflow velocity, temperature,
and density as a function of $z$ for Model 10CW and Model 10C in QW96.
The inner boundary conditions in Model 10C in QW96 is same as those
in Model 10CW (see table 1 in \cite{qian96}). 
From this figure and Eq.~(\ref{eqn15}),
we can see the tendency mentioned above clearly. 
We consider the reason for this tendency below.
In order to
find a clue to it, we show the absolute values (in cgs units) of the
components in Eqs.~(\ref{eqn11})
and~(\ref{eqn12}) for Model 10CW in Figure~\ref{fig2}. From this figure,
we can see what determines the gradients of temperature and density. 
For comparison, we also show those for Model 10C in QW96 in
Figure~\ref{fig3}. Lines (a)-(g) correspond to $4\epsilon/T$,
$(P/\epsilon \rho)(\epsilon+P/ \rho)$, $v_z^2$, $\rho GM/z^2$,
$P\dot{q}/v_z \epsilon$, $(\epsilon/\rho + P/\rho^2)\frac{d \rho}{d z}$,
and $\dot{q}/v_z$ as a function of $z$, respectively.
As can be seen from Figures~\ref{fig2} and~\ref{fig3},
$d \rho / d z$ can be approximated by
\begin{eqnarray}
\frac{d \rho}{d z} \sim - \frac{\frac{\rho GM}{z^2}}{\frac{P}{\epsilon \rho}
\left ( \epsilon + \frac{P}{\rho} \right )  }.
\label{eqn17}
\end{eqnarray}
Also, $d T/ d z$ can be approximated by
\begin{eqnarray}
\frac{d T}{d z} \sim \frac{ \left( \frac{\epsilon}{\rho}
+ \frac{P}{\rho^2} \right) \frac{d \rho}{d z}}{\frac{4 \epsilon}{T}}
\sim -\frac{m_N}{S} \frac{GM}{z^2}.
\label{eqn18}
\end{eqnarray}
From Eq.~(\ref{eqn18}) we can see that the temperature gradient
$d T/ d z$ is steeper when the entropy per baryon is
smaller. So we can expect that the entropy per baryon is larger
in Model 10CW than Model 10C in QW96. It is verified in Figure~\ref{fig4}.
So, the problem why $T_b$ becomes high is translated to the problem
why the entropy per baryon becomes high in Model 10CW. The answer
is shown in Figure~\ref{fig2}. The velocity at small radii in Model 10CW
is smaller than that in Model 10C in QW96. The entropy per baryon
is getting larger with time as long as the heating process is effective.
The heating process is effective only at small radii, which is
proved in Figure~\ref{fig4}. So, if the velocity at small radii
is small, there is a long time for the outflow matter to get
heat from neutrinos. As a result, higher entropy per baryon can be
achieved and $T_b$ becomes higher. This is the case with the jet models
studied in this paper. As for the reason why the initial velocity has to
be smaller in Model 10CW than Model 10C in QW96 is relatively difficult.
At least, we can say that $d v_z/ dz$ can be written from Eq.~(\ref{eqn121})
as
\begin{eqnarray}
\frac{dv_z}{dz} = -\frac{v_z}{\rho}\frac{d \rho}{d z} \;\;\; \rm in \;
Model \; 10CW.
\label{eqn19}
\end{eqnarray}
On the other hand, it can be written as
\begin{eqnarray}
\frac{dv_z}{dz} = -\frac{v_z}{\rho}\frac{d \rho}{d z} - \frac{v_z}{r}
\;\;\; \rm in \; Model \; 10C \; in \; QW96,
\label{eqn20}
\end{eqnarray}
because $4 \pi z^2 \rho v_z$ is constant in Model 10C in QW96.
Taking these equations and Eq.~(\ref{eqn11}), which also holds in
the models in QW96 (\cite{qian96}), into consideration,
the velocity gradient is steeper in
Model 10CW as long as $v_z(z)$, $\rho (z)$, and $T (z)$
are same in both models.
So we can guess that initial $v_z$ has to be smaller in Model 10CW
so as not to achieve the adiabatic sound speed,
$v_s$ = $(4P/3\rho)^{\frac{1}{2}}$. This will be the reason why
the initial velocity in Model 10CW has to be smaller.

\placefigure{fig1}
\placefigure{fig2}
\placefigure{fig3}
\placefigure{fig4}

Next, we consider the dynamical timescale.
As mentioned above, the dynamical timescale becomes infinity when
$T_b$ is above 0.2 MeV. Since $T_b$ in the jet models tends to
be higher, we can not determine the dynamical timescale in many
jet models. This is a bad tendency to produce r-process nuclei.
From Table~\ref{tab2}, the dynamical timescale can be determined
as long as $L_{\bar{\nu}_e, \rm 51}$ is small. We will discuss later
whether r-process can occur or not in such models.
Before we continue to further discussion, we explain the
way to derive the analytically estimated dynamical timescale
$\tau_{\rm dyn,ana}$.

Like Model 10C in QW96, from Eqs.~(\ref{eqn5}) and~(\ref{eqn9}) we
can derive the following equation as long as $\Omega D$ is small:
\begin{eqnarray}
\dot{q} = v_z \frac{d}{dz} \left( \frac{v_z^2}{2} 
+ \frac{TS}{m_N} -\frac{GM}{z}
\right).
\label{eqn21}
\end{eqnarray}
So, when the heating/cooling process becomes non-effective,
there is a conserved quantum
\begin{eqnarray}
\epsilon_{\rm flow, f} = \left( \frac{v_z^2}{2} + \frac{TS}{m_N} 
- \frac{GM}{z}      \right ).
\label{eqn22}
\end{eqnarray}
In the subcritical case, $v_z \le v_s$ everywhere, we can approximate
Eq.~(\ref{eqn22}) as
\begin{eqnarray}
\frac{TS}{m_N} - \frac{GM}{r} \sim \frac{T_b S}{m_N}. 
\label{eqn23}
\end{eqnarray} 
When $T_b \ll$0.5MeV, we can take
\begin{eqnarray}
\frac{TS}{m_N} \sim \frac{GM}{r} \; \rm at \; 0.5 MeV.
\label{eqn24}
\end{eqnarray} 
See Qian and Woosley (1996) for details. 
In this case, $T \propto r^{-1}$ near $T \sim$ 0.5 MeV because
$S$ is also a conserved quantum
as long as the heating/cooling process does not work. From Eq.~(\ref{eqn15}),
$\rho \propto r^{-3}$. From Eq.~(\ref{eqn121}), $v_z \propto r^3$.
So, $\tau_{\rm dyn,ana}$ can be approximated as
\begin{eqnarray}
\tau_{\rm dyn,ana} && = \int^{T=0.2 \rm MeV}_{T=0.5 \rm MeV} dt = 
\int^{T=0.2 \rm MeV}_{T=0.5 \rm MeV} \frac{dt}{dz} \frac{dz}{dT} dT
\\ 
&& = \left ( \frac{r}{v T^2} \right )_{T = 0.5 \rm MeV}
\int^{T=0.5 \rm MeV}_{T=0.2 \rm MeV} T dT \sim 0.42 \left( \frac{r}{v}
\right)_{T = 0.5 \rm MeV} \;\;\; \left [   s   \right ]  .
\label{eqn25}
\end{eqnarray}

For further discussion, we try to express the dynamical timescale
by input parameters (see \cite{qian96} for details).
From simple calculations, we can derive the entropy per baryon
and the mass outflow rate in Model 10CW as
\begin{eqnarray}
S/k \sim 235 L^{-\frac{1}{6}}_{\bar{\nu}_e, \rm 51}
\epsilon^{-\frac{1}{3}}_{\bar{\nu}_e, \rm MeV} 
R^{-\frac{2}{3}}_{\rm 6}
\left(   \frac{M}{1.4M_{\odot}}       \right )
\label{eqn26}
\end{eqnarray} 
and
\begin{eqnarray}
\dot{M} \sim
\frac{D^2}{4R^2} \times
1.14 \times 10^{-10}
L^{\frac{5}{3}}_{\bar{\nu}_e, \rm 51}
\epsilon^{\frac{10}{3}}_{\bar{\nu}_e, \rm MeV} 
R^{\frac{5}{3}}_{\rm 6}
\left(   \frac{1.4 M_{\odot}}{M}       \right )^2 
\;\;\;  \left [   M_{\odot} \; s^{-1}   \right ],
\label{eqn27}
\end{eqnarray} 
respectively.
At last, we can derive the expression for $\tau_{\rm dyn,ana}$ from
Eqs.~(\ref{eqn121}),~(\ref{eqn15}),~(\ref{eqn24}),~(\ref{eqn25}),
~(\ref{eqn26}), and~(\ref{eqn27}) that
\begin{eqnarray}
\tau_{\rm dyn,ana} \sim
10.5
L^{-\frac{4}{3}}_{\bar{\nu}_e, \rm 51}
\epsilon^{-\frac{8}{3}}_{\bar{\nu}_e, \rm MeV} 
R^{\frac{5}{3}}_{\rm 6}
\left(   \frac{M}{1.4M_{\odot}}       \right ) 
\;\;\;  \left [ s   \right ]. 
\label{eqn28}
\end{eqnarray} 
This value for each model is written in Table~\ref{tab2}.
We can find that $\tau_{\rm dyn,ana}$ agrees with $\tau_{\rm dyn}$
within a factor of 2-3 in all models in which the dynamical
timescale can be defined, although $\tau_{\rm dyn,ana}$ tends to be
smaller. This precision of $\tau_{\rm dyn,ana}$ is about the same
that in the analysis in QW96 (see \cite{qian96} for details).

We compare $\tau_{\rm dyn,ana}$ derived here with that in QW96
in order to see the effects of jet. It is written as
\begin{eqnarray}
\frac{\tau_{\rm dyn,ana}}{\tau_{\rm dyn,QW}} = 0.15
\frac{R_6^{\frac{2}{3}}}{L^{\frac{1}{3}}_{\bar{\nu}_e, \rm 51} 
\epsilon^{\frac{2}{3}}_{\bar{\nu}_e, \rm MeV}
 }.
\label{eqn29}
\end{eqnarray} 
For example, when $L_{\bar{\nu}_e, \rm 51}$ = 0.1, $R_6$ = 1,
and $\epsilon_{\bar{\nu}_e, \rm MeV}$ = 22, $\tau_{\rm dyn}/\tau_{\rm dyn,QW}$
= 0.04. Even if we take the uncertainty (a factor of 2-3)
of $\tau_{\rm dyn,ana}$ into consideration, it seems that 
short dynamical timescale can be realized in jet models.
So, as long as $T_b$ can be small (that is, $L_{\bar{\nu}_e, \rm 51}$
is small enough), we think that
short dynamical timescale can be realized in jet models.

Now we discuss whether r-process nucleosynthesis can occur in jet models
or not. We can use the Hoffman's criterion (\cite{hoffman97}) for the
judgment. The Hoffman's criterion can be written as
\begin{eqnarray}
S \ge 2 \times 10^3 Y_e \left( \frac{\tau_{\rm dyn}}{s}
\right )^{\frac{1}{3}}
\label{eqn30}
\end{eqnarray}
for $Y_e \ge$0.38. This is the criterion for production of r-process nuclei
with mass number $A \sim$200 (\cite{hoffman97}).
Substituting Eqs.~(\ref{eqn26}) and~(\ref{eqn28}) into Eq.~(\ref{eqn30}),
We can translate the criterion into
\begin{eqnarray}
R_6 \le \frac{0.74}{C^{\frac{3}{11}}} L_{\bar{\nu}_e, \rm 51}^{\frac{5}{22}}
\left(  \frac{M}{1.4 M_{\odot}}  
\right )^{\frac{6}{11}},
\label{eqn31}
\end{eqnarray}
where $\epsilon_{\bar{\nu}_e, \rm MeV}$ is assumed to be 22.
$C$ is the factor 2-3 which represents the tendency
that $\tau_{\rm dyn,ana}$ tends to be smaller than $\tau_{\rm dyn}$.
In case of $M$=1.4$M_{\odot}$ and $C$=2, we can see
from Eq.~(\ref{eqn31}) that $R_6$ can be larger than 1
as long as $L_{\bar{\nu}_e, \rm 51}$ is larger than 8.6.
This means that there seems to be possible to produce r-process
nuclei without unrealistically soft EOS as long as
$L_{\bar{\nu}_e, \rm 51}$ is larger than 8.6. However,
we found from Table~\ref{tab2} that $T_b$ can not be lower than
0.2 MeV when the neutrino luminosity is so high. So we have to
conclude that r-process can not occur in these models.

However, we can find that the requirement to produce r-process
nuclei is relaxed in these jet models. In fact, for the models
of QW96, the criterion like Eq.~(\ref{eqn31}) can be written as
\begin{eqnarray}
R_{6, \rm QW} \le 0.19 L_{\bar{\nu}_e, \rm 51}^{\frac{1}{6}}
\left(  \frac{M}{1.4 M_{\odot}}  
\right )^{\frac{2}{3}}.
\label{eqn32}
\end{eqnarray}
So, $R_{6,\rm QW}$ can be larger than 1 when $L_{\bar{\nu}_e, \rm 51}$
is larger than 2.3$\times10^{4}$, which can not be realized
in numerical simulations of supernovae (\cite{woosley94}).
We can compare the required radius in jet models ($C=2$)
with that in models of QW96 as
\begin{eqnarray}
\frac{R_6}{R_{6, \rm QW}} = 3.2 L_{\bar{\nu}_e, \rm 51}^{\frac{2}{33}}
\left(  \frac{M}{1.4 M_{\odot}}  \right )^{-\frac{4}{33}}.
\label{eqn33}
\end{eqnarray}
For example, in case of $M$ = 1.4$M_{\odot}$
and $L_{\bar{\nu}_e, \rm 51}$ = 0.1,
$R_6/R_{6, \rm QW}$ = 2.8.
This suggests that the EOS of the nuclear matter has not necessary
to be too soft as Otsuki et al (2000) requires, when the jet models
are adopted. So, we can conclude that the requirement to achieve
r-process nucleosynthesis is relaxed
for the jet models, although r-process nucleosynthesis is not still
able to occur in the jet models. When we include the effects of a jet
and general relativity at the same time, an successful solution
may be obtained. We will discuss this topic in the forthcoming paper.

Finally, we consider the mass outflow rate.
Taking the event rate of
collapse-driven supernovae ($10^{-2} \; \rm yr^{-1} \; Gal^{-1}$;
\cite{vandenbergh91}), one needs $10^{-6}$ to $10^{-4}$ $M_{\odot}$
of ejected r-process material per collapse-driven supernova event
to explain the observed solar-abundance ratio (\cite{kappeler89};
\cite{woosley94}). We give a rough estimate whether each model
can produce r-process matter enough to explain the solar-system
abundance. Taking the gravitational binding energy of a
neutron star is $\sim 3 \times 10^{53}$ erg into consideration,
the duration time, $t$, can be roughly estimated as
\begin{eqnarray}
t \le \frac{300}{6 \times L_{\bar{\nu}_e, \rm 51}} \;\;\; \rm \left [
 s \right ]
\label{eqn34}
\end{eqnarray}
where the factor 6 represents the number of flavor of neutrinos.
So required $\dot{M}_{\rm req}$ to explain the solar abundance ratio
is
\begin{eqnarray}
\dot{M}_{\rm req} \ge \left(10^{-4} - 10^{-6}      \right ) M_{\odot}
\times   \frac{L_{\bar{\nu}_e, \rm 51}}{50} \;\;\;  
\left [ M_{\odot} \; s^{-1}    \right ]     .
\label{eqn35}
\end{eqnarray}
So $\dot{M}$ in Table~\ref{tab1} has to meet the relation
\begin{eqnarray}
2 \times \dot{M} \sim \dot{M}_{\rm req}
\label{eqn36}
\end{eqnarray}
where the factor 2 represents that the matter is ejected from
both (north and south) sides of the polar region.
We can find from Eq.~(\ref{eqn35}),~(\ref{eqn36}), and Table~\ref{tab1}
that low-luminosity models like 10GW-10LW and 30FW-30HW can not
meet the relation mentioned above, although the high-luminosity models
like 10AW-10FW and 30AW-30EW may be able to meet the relation.

At last we can derive the conclusion. In the framework of the steady
neutrino-driven jet with weak magnetic fields, high-luminosity models
can not work because $T_b$ becomes high and dynamical timescale becomes
too long. Also, low-luminosity models do not present a required condition
to produce the r-process nuclei in the framework of Newtonian gravity.
Although a successful solution may be obtained when we include
the effects of jet and general relativity at the same time, the amount
of mass outflow seems to be too little to explain the solar-system
abundance ratio in such low-luminosity models.
As a conclusion, we have to say that it is difficult to cause
a successful r-process nucleosynthesis in the weak field solutions.

\subsection{Strong Field Solutions}\label{strong}
\indent

In this subsection, we examine the models in which $\Omega D \ne$0.
As stated in subsection~\ref{jetmodel}, the truth is that the
solutions studied in this paper are inapplicable to the case
of strong magnetic fields. However, we want to see what can
happen as the magnetic fields are getting stronger and stronger.
At the same time, we should investigate the range of application
of weak field solutions.

In this study, we consider three models (Model 10GS1-3) for strong
field solutions.
These are listed in Table~\ref{tab1}. The strength of the magnetic
fields corresponds to 2.5$\times 10^{13}$ G, 2.5$\times 10^{14}$ G, and
2.5$\times 10^{15}$ G, respectively (see Eq.~(\ref{eqn10})).

The results are shown in Table~\ref{tab2}. We can find that
the values of output parameters change drastically in Model 10GS3.
This result suggests that the weak field solutions can be applied
at least in the
range of order $B \le 10^{14}$ G. This means that
the weak field solutions will be valid for the typical
proto-neutron stars. In case we consider the nucleosynthesis in a
magnetar (\cite{woods99}), we will have to investigate using
strong field solutions.

\begin{figure}
\epsscale{1.0}
\plotone{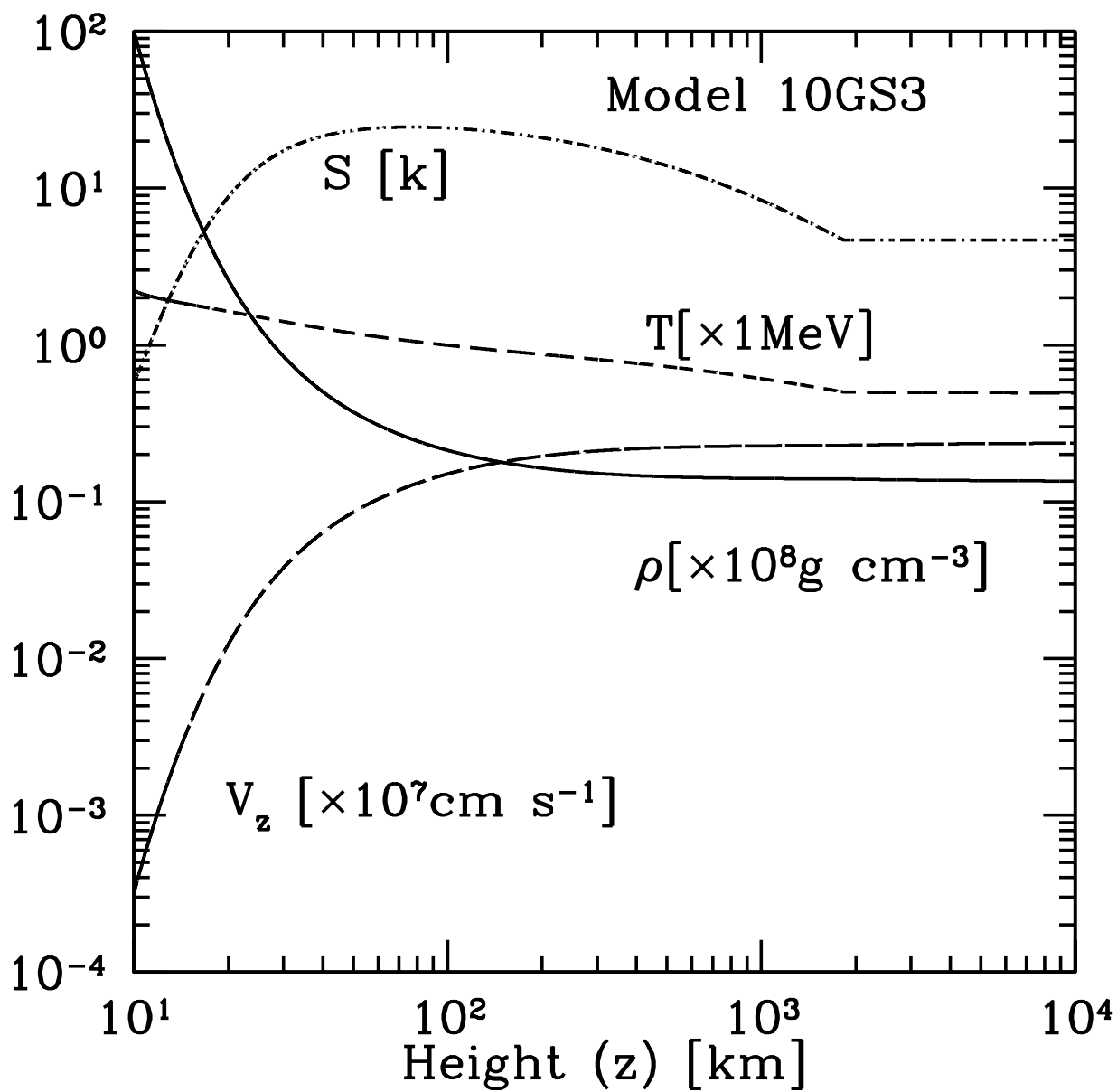}
\figcaption{
Outflow velocity, temperaure, density, and entropy per baryon
as a function of height ($z$) from the center of the proto-neutron star
in Model 10GS3.
These quanta are written in unit of $10^7$ cm $\rm s^{-1}$, 1 MeV,
$10^8$ g cm$^{-3}$, and the boltzman constant, respectively.
\label{fig5}}
\end{figure}

\begin{figure}
\epsscale{1.0}
\plotone{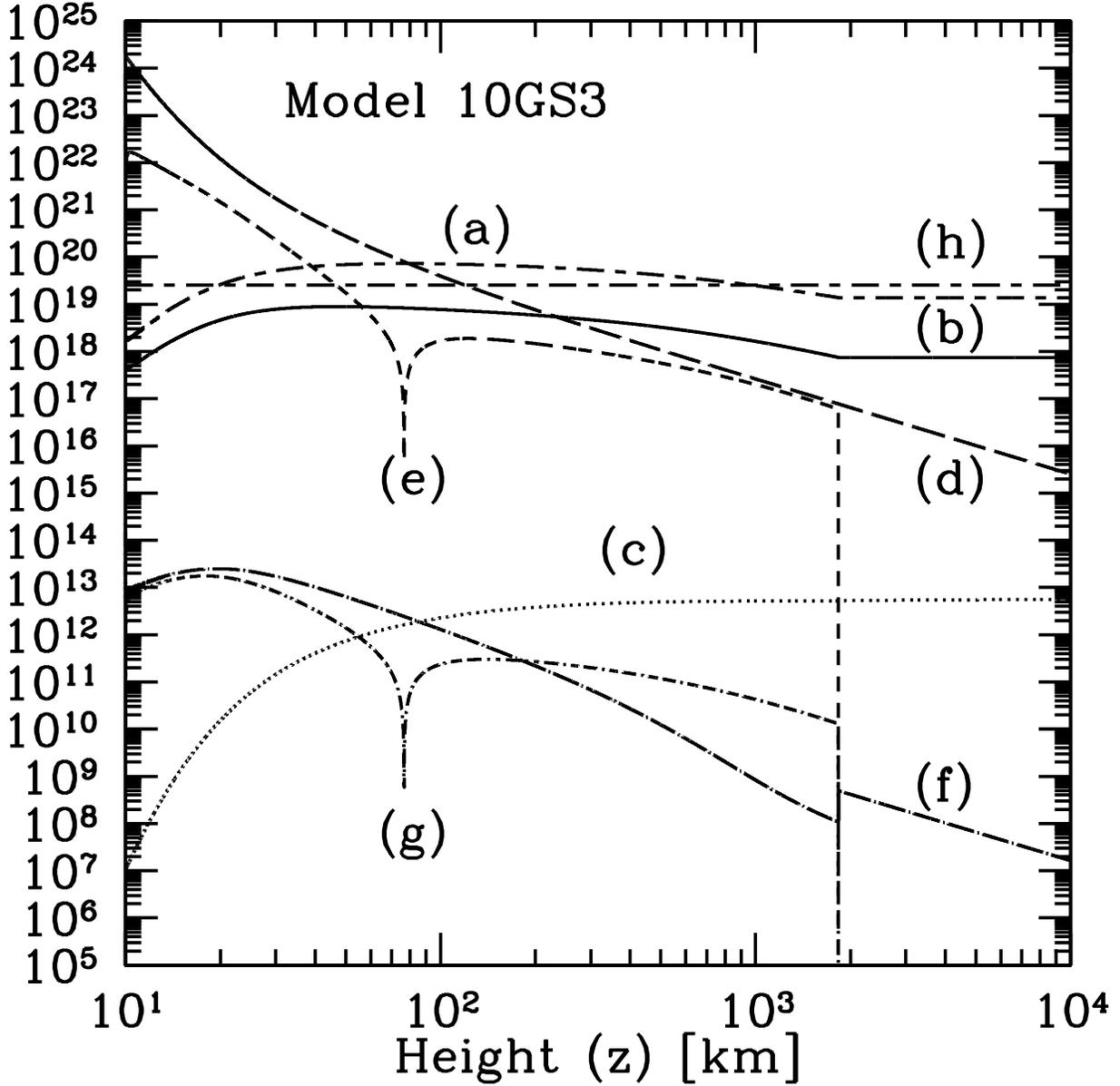}
\figcaption{
Absolute values (in cgs units) of the components of
Eqs.~(16) and~(17) for Model 10GS3.
Lines (a)-(h) correspond to $4\epsilon/T$,
$(P/\epsilon \rho)(\epsilon+P/ \rho)$, $v_z^2$, $\rho GM/z^2$,
$P\dot{q}/v_z \epsilon$, $(\epsilon/\rho + P/\rho^2)\frac{d \rho}{d z}$,
$\dot{q}/v_z$, and $\Omega^2 D^2/4$ as a function of height ($z$) from
the center of the proto-neutron star, respectively.
The discontinuities of lines (e) and (g) at $z \sim 8 \times 10^6$ cm
and at $z \sim 2 \times 10^8$ cm reflect the change of the sign of $\dot{q}$
and freezeout of the neutrino reactions (see subsection 2.1), respectively.
\label{fig6}}
\end{figure}

In order to understand what happens in Model 10GS3, we show the
outflow velocity, temperature, density, and entropy per baryon
as a function of $z$ of
Model 10GS3 in Figure~\ref{fig5}. We also show the absolute values
(in cgs units) of the
components of Eqs.~(\ref{eqn11}) and~(\ref{eqn12}) for Model 10GS3
in Figure~\ref{fig6}. Lines (a)-(h) correspond to $4\epsilon/T$,
$(P/\epsilon \rho)(\epsilon+P/ \rho)$, $v_z^2$, $\rho GM/z^2$,
$P\dot{q}/v_z \epsilon$, $(\epsilon/\rho + P/\rho^2)\frac{d \rho}{d z}$,
$\dot{q}/v_z$, and $\Omega^2 D^2$/4 as a function of $z$, respectively.
As can be seen from Figures~\ref{fig6},
$d \rho / d z$ at $r=D$ can be approximated by
\begin{eqnarray}
\left.  \frac{d \rho}{d z} \right|_{r=D}    \sim 
-\frac{\rho GM/z^2}{\Omega^2 D^2/4}.
\label{eqn37}
\end{eqnarray}
Also, $d T/ d z$ can be approximated by
\begin{eqnarray}
\frac{d T}{d z} \sim    \frac{ T \left( \frac{\epsilon}{\rho}
+ \frac{P}{\rho^2} \right) \frac{d \rho}{d z}}{4 \epsilon} &&
\;\;\; \rm for \;\; {\it z} \le 8 \times 10^1 \;\; and \;\; {\it z} 
\ge 2 \times 10^3 \\
\sim   \frac{T \dot{q}}{4 \epsilon v_z} && \;\;\; \rm for \;\;
8 \times 10^1   \le {\it z} \le 2 \times 10^3.
\label{eqn38}
\end{eqnarray}

\placefigure{fig5}
\placefigure{fig6}

We note that the denominator of Eq.~(\ref{eqn37}) is $\Omega^2 D^2$/4,
which is constant through out. Since the denominator is large, the value
of the density gradient becomes small. So the density gradient in 10GS3
becomes nearly zero
at smaller $z$ (compare Figures~\ref{fig1} and~\ref{fig5}).
As a result, temperature does not decrease due to the adiabatic expansion.
When the temperature is kept to be high,
cooling process works better (see Eqs.~(\ref{eqn52}),~(\ref{eqn53}),
and~(\ref{eqn54})). So, at $z \sim 8 
\times 10^6$ cm, the cooling process overcomes the heating process.
As a result, entropy per baryon is getting lower in the range
$8 \times 10^6  \le z \le 2 \times 10^8$ cm. This is the reason
why the entropy per baryon becomes small in Model 10GS3.
We also note that the temperature gradient becomes too small
at $z \ge 2 \times 10^8$ cm. This is because the numerator
of Eq.~(44) becomes in proportion to $d \rho / dz$, which
is very small as seen above. So we can find that the entropy per
baryon is small and the dynamical timescale is long in the
strong field solution 10GS3. This tendency is a bad one for the production
of r-process nuclei. 
The truth is, of course, that the solutions studied in this paper
are inapplicable to the case of
strong magnetic fields (see Eq.~(\ref{eqn11})).
However, similar tendency discussed here may be also found in
the exact solutions. If so,
the strong field solutions are unsuitable for the
r-process nucleosynthesis.

Finally we add a comment on the effects of the magnetic fields.
You may have thought that the dynamical timescale in the strong
field solutions will become shorter than that in the weak field solutions,
because the magnetic fields accelerate
the matter as long as the density gradient is negative (see
Eqs.~(\ref{eqn9}) and~(\ref{eqn10})).
However, in the framework of the steady flow, the density
gradient becomes nearly zero and acceleration does not occur.
So, the situation may be different and acceleration occurs
in the case of an unsteady flow. We will investigate the features
of such an unsteady flow by numerical tests in the near future.

\section{SUMMARY AND DISCUSSION} \label{summary}
\indent

We have studied whether the jet in a collapse-driven
supernova can be a key process for the r-process nucleosynthesis.
We have studied the effects of a jet using a simple model, because
the results of a realistic numerical simulation concerning with such a theme
will not be understood completely without analytical study.
Although our final goal is to perform such realistic numerical simulations,
this work is a necessary process to understand the effects of a jet
on the r-process nucleosynthesis.

We have studied two cases, that is, the cases in which the magnetic
fields are weak/strong. In both cases, we assumed that the flow is
steady and subsonic.

As for the weak field solutions, we have concluded that high-luminosity models
can not work because $T_b$ becomes high and dynamical timescale becomes
too long.
Also, low-luminosity models do not present a required condition
to produce the r-process nuclei in the framework of Newtonian gravity.
Although a successful solution may be obtained when we include
the effects of jet and general relativity at the same time, the
total mass
of the outflow seems to be too little to explain the solar-system
abundance ratio in such low-luminosity models.
On the other hand, we found that the entropy per baryon is small
and the dynamical timescale is long in the strong field solution.
We can tell that this tendency is a bad one for the production
of r-process nuclei. 
The truth is that the solutions studied in this paper
are inapplicable to the case of strong magnetic fields.
However, similar tendency may be also found in
the exact solutions. If so, we will be able to conclude that
strong field solutions are unsuitable for r-process nucleosynthesis.
As a conclusion, we have to say that it seems to be difficult to cause
a successful r-process nucleosynthesis in the jet models in this study.

We have to emphasize that there are some assumptions in this study.
So, we can not say that we have proved that a successful r-process
nucleosynthesis does not occur in a neutrino-driven jet
in a collapse-driven supernova. For example, we assumed that the
flow is subsonic and there is no critical point,
which is the common assumption in the previous studies
on the r-process nucleosynthesis (\cite{qian96}; \cite{otsuki00}).
However, we think that we do not need to restrict the solutions
in such a way, that is, there may be a transition point
at which Eqs.~(\ref{eqn5}),~(\ref{eqn9}),~(\ref{eqn11}), and~(\ref{eqn12})
break down. In most cases, the transition
point will be a shock front. It means that the flow will gain entropy
at the transition point, which will be a good sense to produce
the r-process nuclei.
The problem whether the flow contains transition points or not
depends sensitively on the initial velocity on the surface
of the proto-neutron star. So our final goal is to determine physically
the velocity at the surface of the proto-neutron star.
It means that the $\dot{M}$ should be not given as an input parameter.
It should be an output parameter. We have to investigate the mechanism
for determining the outflow velocity at the surface of the proto-neutron
star for further discussions. We also assumed that the flow is steady.
As seen in subsection~\ref{strong}, the density
gradient becomes nearly zero and acceleration due to the
strong magnetic fields does not occur in the framework of the steady flow.
So, the situation may be different and acceleration occurs
in the case of the unsteady flow.
We should investigate the features of the unsteady jet flow.
It will be investigated
by numerical tests assuming a simple environment. We will perform
such numerical tests in the near future.
At the same time, we will try to perform realistic numerical
simulations for the jet induced explosion in a collapse-driven
supernova and seek a final answer whether the jet in a collapse-driven
supernova is a key process for the r-process nucleosynthesis or not.

\acknowledgements
This research has been supported in part by a Grant-in-Aid for the
Center-of-Excellence (COE) Research (07CE2002) and for the Scientific
Research Fund (199908802) of the Ministry of Education, Science, Sports and
Culture in Japan and by Japan Society for the Promotion of Science
Postdoctoral Fellowships for Research Abroad.

\begin{table*}
\begin{center}
\begin{tabular}{ccccllccccccccccc}
\tableline
\tableline
      & Mass & Radius & $L_{\bar{\nu}_e}$ & $\dot{M}$ & $\Omega D$ \\ 
Model & ($M_{\odot}$) & (km)          & ($10^{51}$ ergs $\rm s^{-1}$) 
      & ($M_{\odot}$ $\rm s^{-1}$)    & (cm $\rm s^{-1}$) \\
\tableline
10AW & 1.4 & 10 & 3.00  & 8.4(-8) & 0 \\
10BW & 1.4 & 10 & 1.00  & 1.4(-8) & 0 \\
10CW & 1.4 & 10 & 0.60  & 5.8(-9) & 0 \\
10DW & 2.0 & 10 & 3.00  & 5.1(-8) & 0 \\
10EW & 2.0 & 10 & 1.00  & 8.1(-9) & 0 \\
10FW & 2.0 & 10 & 0.60  & 3.5(-9) & 0 \\
10GW & 1.4 & 10 & 0.10  & 3.1(-10)& 0 \\
10GS1& 1.4 & 10 & 0.10  & 3.1(-10)& 1.0(+8) \\
10GS2& 1.4 & 10 & 0.10  & 3.3(-10)& 1.0(+9) \\
10GS3& 1.4 & 10 & 0.10  & 5.1(-10)& 1.0(+10) \\
10HW & 1.4 & 10 & 0.05  & 9.8(-11)& 0 \\
10IW & 1.4 & 10 & 0.01  & 6.6(-12)& 0 \\
10JW & 2.0 & 10 & 0.10  & 1.8(-10)& 0 \\
10KW & 2.0 & 10 & 0.05  & 5.7(-11)& 0 \\
10LW & 2.0 & 10 & 0.01  & 3.9(-12)& 0 \\
30AW & 1.4 & 30 & 30.0  & 1.3(-6) & 0 \\
30BW & 1.4 & 30 & 10.0  & 2.2(-7) & 0 \\
30CW & 1.4 & 30 & 6.00  & 1.1(-7) & 0 \\
30DW & 1.4 & 30 & 1.00  & 6.0(-9) & 0 \\
30EW & 1.4 & 30 & 0.50  & 1.9(-9) & 0 \\
30FW & 1.4 & 30 & 0.10  & 1.4(-10)& 0 \\
30GW & 1.4 & 30 & 0.05  & 4.4(-11)& 0 \\
30HW & 1.4 & 30 & 0.01  & 3.0(-12)& 0 \\
\tableline
\end{tabular}
\tablenum{1}
\caption{
Model names and input parameters. Mass and radius of the neutron star,
total luminosity of neutrinos, mass outflow rate, and $\Omega D$ are
showm respectively.
}\label{tab1}
\end{center}
\end{table*}

\begin{table*}
\begin{center}
\begin{tabular}{ccllclccccccccccc}
\tableline
\tableline
      & $S$     & $\tau_{\rm dyn}$ & $\tau_{\rm dyn, ana}$ & $Y_e$ & $T_b$  \\ 
Model & ($k$) & (s)              & (s)                &    & MeV            \\
\tableline
10AW & 92       & ---------        & ---------             & 0.43  & 0.68   \\
10BW & 102      & ---------        & ---------             & 0.43  & 0.37   \\
10CW & 110      & ---------        & ---------             & 0.43  & 0.36   \\
10DW & 129      & ---------        & ---------             & 0.43  & 0.62   \\
10EW & 146      & ---------        & ---------             & 0.43  & 0.38   \\
10FW & 156      & ---------        & ---------             & 0.43  & 0.32   \\
10GW & 136      & 1.1(-1)          & 4.1(-2)               & 0.43  & 0.14   \\
10GS1& 137      & 1.1(-1)          & undefined             & 0.43  & 0.15   \\
10GS2& 132      & 1.1(-1)          & undefined             & 0.43  & 0.15   \\
10GS3& 4.7      & ---------        & undefined             & 0.43  & 0.50   \\
10HW & 150      & 1.7(-1)          & 9.4(-2)               & 0.43  & 0.11   \\
10IW & 196      & 1.2(-0)          & 7.8(-1)               & 0.43  & 9.2(-2)\\
10JW & 196      & 1.2(-1)          & 4.8(-2)               & 0.43  & 0.12   \\
10KW & 216      & 2.0(-1)          & 1.1(-1)               & 0.43  & 0.11   \\
10LW & 274      & 1.2(-0)          & 8.9(-1)               & 0.43  & 6.1(-2)\\
30AW & 37       & ---------        & ---------             & 0.43  & 1.21   \\
30BW & 42       & ---------        & ---------             & 0.43  & 0.92   \\
30CW & 42       & ---------        & ---------             & 0.43  & 0.60   \\
30DW & 50       & ---------        & ---------             & 0.43  & 0.31   \\
30EW & 55       & ---------        & ---------             & 0.43  & 0.28   \\
30FW & 64       & 6.5(-1)          & 3.6(-1)               & 0.43  & 0.11   \\
30GW & 72       & 1.6(-0)          & 9.0(-1)               & 0.43  & 0.11   \\
30HW & 90       & 1.1(+1)          & 7.9(-0)               & 0.43  & 6.8(-2)\\
\tableline
\end{tabular}
\tablenum{2}
\caption{
Model names and output parameters.
Entropy per baryon, dynamical timescase ($\tau_{\rm dyn}$),
analytically estimated dynamical timescale ($\tau_{\rm dyn.ana}$),
electron fraction, final temperature are shown respectively.
The reason why dynamical timescales are not written in some models
is that temperature does not decrease to 0.2 MeV within $z$ = $10^9$ cm.
}\label{tab2}
\end{center}
\end{table*}

\end{document}